\documentclass[11pt,a4paper]{article}
\usepackage{amsfonts}
\usepackage{amsmath}
\usepackage{array}
\usepackage{calc}
\usepackage{array}
\usepackage{indentfirst}
\usepackage{graphicx}
\usepackage{hyperref}

\usepackage[svgnames,table]{xcolor}
\usepackage{colortbl}
\newlength\celldim 
\newlength\fontheight 
\newlength\extraheight
\newcounter{sqcolumns}

\newcommand\nl{\tabularnewline\hline}

\newcolumntype{S}
{@{}
>{\centering \rule[-0.5\extraheight]{0pt}{\fontheight + \extraheight}} 
p{\celldim} @{} }
\newcolumntype{Z}{ @{} >{\centering} p{\celldim} @{}}

\arrayrulecolor{White}  
\newcommand{\ka}{\cellcolor{Grey!15}} 
\newcommand{\kb}{\cellcolor{Grey!40}}  

\newcommand{\ko}{\cellcolor{Grey!8}} 
\newcommand{\kx}{\cellcolor{Salmon}} 
\newcommand{\ky}{\cellcolor{LightSkyBlue}} 
\newcommand{\kw}{\cellcolor{Yellow!55}} 
\newcommand{\kv}{\cellcolor{YellowGreen}} 
\newcommand{\ku}{\cellcolor{Orange!70}} 
\newcommand{\kq}{\cellcolor{Cyan!19!Yellow}} 
\newcommand{\ks}{\cellcolor{Beige}} 

\newenvironment{squarecells}[1]
{
\arrayrulewidth=3pt
\setlength\celldim{2.1em} 
\settoheight\fontheight{A}
\setlength\extraheight{\celldim - \fontheight}
\setcounter{sqcolumns}{#1 - 1}
\begin{tabular}{|S|*{\value{sqcolumns}}{Z|}}\hline
 }
{
\end{tabular}
}

\begin{document}
\title{Resonant algebras and gravity}
\author{R. Durka$^{1}$%
\thanks{remigiuszdurka@gmail.com},\\{\small $^{2}$\textit{Instituto de F\'{\i}sica, Pontificia Universidad
Cat\'{o}lica de Valpara\'{\i}so,} }\\{\small Casilla 4059, Valpara\'{\i}so, Chile}}

\maketitle

\begin{abstract}

The $S$-expansion framework is analyzed in the context of a freedom in closing the multiplication tables for the abelian semigroups. Including the possibility of the zero element in the resonant decomposition and associating the Lorentz generator with the semigroup identity element leads to the wide class of the expanded Lie algebras introducing interesting modifications to the gauge gravity theories. Among the results, we find all the Maxwell algebras of type $\mathfrak{B}_m$, $\mathfrak{C}_m$, and recently introduced $\mathfrak{D}_m$. The additional new examples complete resulting generalization of the bosonic enlargements to an arbitrary number of the Lorentz-like and translational-like generators. Some further prospects concerning enlarging the algebras are discussed, along with providing all the necessary constituents for constructing the gravity actions based on the obtained results.
\\[0.82em]
{\it keywords:} resonant algebras, S-expansion, semigroup expansion, Maxwell algebra

\end{abstract}

\section{Introduction}

Following the recent introduction in Ref.~\cite{Concha:2016hbt} of the Maxwell-like $\mathfrak{D}_m$ family of algebras, a question has been raised of finding other examples, which may be interesting in the context of gravity or supergravity. The so-called $S$-expansion procedure \cite{Izaurieta:2006zz,Salgado:2014qqa,Diaz:2012zza} by the use of the abelian semigroups introduces a very general and quite convenient description for this kind of algebraic enlargement. Starting from the Anti-de Sitter (AdS) algebra, one can reproduce the Maxwell algebra \cite{Schrader:1972zd,Bacry:1970ye} introduced in the 70's and the algebra proposed by Soroka-Soroka \cite{Soroka:2004fj, Soroka:2006aj}, and generalize them into two Maxwell families \cite{Salgado:2014qqa,Diaz:2012zza}, which following the notation of Ref.~\cite{Concha:2016hbt}, we denote as $\mathfrak{B}_m$ and $\mathfrak{C}_m$ with subindex indicating $(m-1)$ different types of the generators. A new $\mathfrak{D}_m$ family can be also casted into this scheme, but the way to achieve this was not straightforward and only the presence of the direct sums in
these algebras allowed for generalization to an arbitrary number~$m$.

As we will see, all these families (and many other examples) can arise quite naturally by concerning the freedom in closing the algebras. They will be achieved through closing the particular semigroups obeying the so-called resonant decomposition, but now with possibility of a zero element along with the semigroup identity element associated with the Lorentz generator.

To some extent it could be possible to find the new algebras at the level of generators, simply by completing the commutators to fulfill the Jacobi identities. Eventually however, to construct the gravity actions it would be still necessary to make a transition to the $S$-expansion to obtain the invariant tensors. Naturally, from the semigroup multiplication tables one can easily read off the explicit algebra of the generators (see Appendix~\ref{AppA}). Moreover, such a framework allows us to present the resulting schemes in a very compact and illustrative way. 

The work is organized as follows: we start with the standard $S$-expansion setup and then modify some of the conditions. In Section 3 we reproduce all known Maxwell algebraic families and obtain a general scheme for the new ones. Sections 4 and 5 will be devoted to a discussion of the non-standardly enlarged algebras, whereas Section 6 will offer details concerning construction of the gravity actions.

\section{Resonant decomposition with zero and identity elements}

Suitable conditions required to provide reasonable grounds for the theories of gravity constructed as the gauge theories begin with the commutator of the Lorentz generator  $\tilde{J}_{ab}$ with itself and the generator of translations $\tilde{P}_{a}$. This becomes essential in gauging the algebra because to obtain the field transformations and construct the curvature two-form $F=dA+A\wedge A=\frac{1}{2}F_{\mu\nu}^{M}(x)\mathbb{T}_{M}\,dx^{\mu}\wedge
dx^{\nu}$, for a given gauge parameter $\Theta=\Theta^M(x)\mathbb{T}_{M}$ and a connection one-form $A=A_{\mu}^{M}(x)\mathbb{T}_{M}\,dx^{\mu}$, one needs to evaluate 
\begin{align}
\delta _{\Theta }A_{\mu }& =-(\partial _{\mu }\Theta +[A%
_{\mu },\Theta ]) \qquad\text{and}\qquad F_{\mu\nu}=\partial_{\mu}A_{\nu}-\partial_{\nu}A_{\mu}%
+[A_{\mu},A_{\nu}]\,.
\end{align}
Indeed, for the AdS-valued connection, $A=\frac{1}{2}\omega^{ab}\tilde{J}_{ab}+\frac{1}{\ell}e^a \tilde{P}_a$ with group indices $a,b=1,...,D$, the Riemann curvature and torsion two-forms are associated with the following commutators
\begin{align}
\left[  \tilde{J}_{ab},\tilde{J}_{cd}\right]   &  =\eta_{bc}\tilde{J}%
_{ad}-\eta_{ac}\tilde{J}_{bd}-\eta_{bd}\tilde{J}_{ac}+\eta_{ad}\tilde{J}%
_{bc}\, &\to\qquad &  R^{ab} =d\omega^{ab}+\omega^{ac}\wedge\omega_c^{~b}\,,\label{equ1}\\
\left[  \tilde{J}_{ab},\tilde{P}_{c}\right]   &  =\eta_{bc}\tilde{P}_{a}%
-\eta_{ac}\tilde{P}_{b}  & \to \qquad & T^a=de^a+\omega^a_{~b}\wedge e^b\,.\label{equ2}
\end{align}
Altogether with the last piece,
\begin{align}
\left[  \tilde{P}_{a},\tilde{P}_{b}\right]   =0\qquad \text{or}\qquad \left[  \tilde{P}_{a},\tilde{P}_{b}\right]  =\tilde{J}_{ab}\qquad \text{or} \qquad \left[  \tilde{P}_{a},\tilde{P}_{b}\right]=-\tilde{J}_{ab}\,,\label{equ3}
\end{align}
this obviously corresponds to the Poincar\'{e} ($ISO$), Anti de-Sitter (AdS) and de-Sitter (dS) group of symmetries, which brings the contribution of zero or $\pm \frac{1}{\ell^2}e^a\wedge e^b$ to the Lorentz part of curvature. The conventions used throughout this paper, due to possible future supergravity applications, will account only for the first two cases (Poincar\'{e} and AdS).

In the $S$-expansion method \cite{Izaurieta:2006zz,Salgado:2014qqa} the new algebras are derived from the AdS by using a
particular choice of the abelian semigroup. The procedure starts from the
decomposition of the original algebra $\mathfrak{g}$ into subspaces,
\begin{equation}
\mathfrak{g}=\mathfrak{so}\left(  D-1,2\right)  =\mathfrak{so}\left(
D-1,1\right)  \oplus\frac{\mathfrak{so}\left(  D-1,2\right)  }{\mathfrak{so}%
\left(  D-1,1\right)  }=V_{0}\oplus V_{1}\,,
\end{equation}
where $V_{0}$ is spanned by the Lorentz generator $\tilde{J}_{ab}$ and $V_{1}$
by the AdS translation generator $\tilde{P}_{a}$. The subspace structure can be then written as
\begin{equation}
\left[  V_{0},V_{0}\right]  \subset V_{0}\,,\qquad\left[  V_{0},V_{1}\right]
\subset V_{1}\,,\qquad\left[  V_{1},V_{1}\right]  \subset V_{0}\,. \label{subspace}%
\end{equation}
If we define $S=\left\{\lambda_{0},\lambda_{1},\ldots\right\}  $ as an abelian semigroup with the multiplication law being associative and commutative, then the Lie algebra $\mathfrak{G} = S\times \mathfrak{g}$ is called the $S$-expanded algebra of $\mathfrak{g}$. For the \textit{resonant subset decomposition} $S=S_{0}\cup
S_{1}\,$, with%
\begin{align}
S_{0}  =\left\{ \lambda_{2i}\right\}\qquad\text{and}\qquad S_{1} =\left\{ \lambda_{2j+1}\right\}\,\quad\text{ \ for }i,j=0,1,2,\ldots\,,
\end{align}
the new algebra will be spanned by the $\left\{  J_{ab,\left(  i\right)
},P_{a,\left(  j\right)  }\right\}  $, where the new generators are related to the original $\mathfrak{so}\left(  D-1,2\right)  $ ones through%
\begin{align}
J_{ab,\left(  i\right)  }=\lambda_{2i}\tilde{J}_{ab}\text{\qquad and \qquad}P_{a,\left(  j\right)  }=\lambda_{2j+1}\tilde{P}_{a}\,.\label{new-generators}
\end{align}
This decomposition satisfies
\begin{equation}
S_{0}\cdot S_{0}\subset S_{0}\,,\qquad S_{0}\cdot S_{1}\subset S_{1}\,,\qquad
S_{1}\cdot S_{1}\subset S_{0}\,.\label{resonant}
\end{equation}
and was called \textit{resonant} since it has the same structure as (\ref{subspace}). The final form of the algebra will depend on the number of elements and particular law of semigroup multiplication. Such a setup for $S=\{\lambda_\alpha\}^{N}_{\alpha=0}$ with the algebra subindex $m=N+2$, a priori does not contain the zero element. One needs to include it as $\lambda_{N+1}$ element and apply the so-called $0_S$-reduction process \cite{Izaurieta:2006zz,Salgado:2014qqa,Diaz:2012zza} to finally obtain the Maxwell $\mathfrak{B}_m$ family and make connection with an expansion by the means of the Maurer-Cartan forms \cite{deAzcarraga:2002xi}.

Let us now do things differently and just start with the minimal set of elements with some minimal multiplication rules and then check what the freedom is in closing them to form a semigroup useful in the gravity context. Naturally, to provide the most general treatment, the resonant decomposition from Eq.~(\ref{resonant}) gets extended to the eventuality of having an explicit zero element\footnote{In such situations, though, we will omit presenting the additional row and column corresponding to the $0_S$ element.} in $\bar{S}_{0}=S_0 \cup \left\{0_S\right\}$ and $\bar{S}_{1}=S_1 \cup \left\{0_S\right\}$. This absorbing element, for which $0_S \lambda_k=\lambda_k 0_S=0_S$, acting on a generator corresponds to $0_S \mathbb{T}_{M}=0$.

Explicitly marking the resonant decomposition in the multiplication tables can help us fix the missing entries. Thus, subsequent separation in the outcome of the multiplications will be indicated by \fboxsep=1.mm \fboxrule=0mm \fcolorbox{black}{Gray!15}{$\bar{S}_0$}~and~\fboxsep=1.mm \fboxrule=0mm \fcolorbox{black}{Gray!40}{$\bar{S}_1$} describing, respectively, the elements later associated with the Lorentz-like and translation-like generators. 

There will be, nevertheless, an important exception. We are going to introduce the semigroup identity element, chosen to be $\lambda_0$, and relate it to the Lorentz generator. Therefore, effectively we deal here with the monoids. Mind that then $\lambda_0$ preserves all the elements, so the zero element is excluded from the results of $\lambda_0$ times whatever element other than $0_S$. That alone provides already two starting conditions related to the commutators $[J_{ab},J_{cd}]$ and $[J_{ab},P_{c}]$,
\begin{align}
\lambda_0 \lambda_0& =\lambda_0\,,\\
\lambda_0 \lambda_1& =\lambda_1\,,
\end{align}
which were needed for the right definitions of the curvature and torsion in Eq.~(\ref{equ1})-(\ref{equ2}).

After putting all of that into the commutative multiplication table (but keeping in mind that the $\lambda_0$ row and column come with the restriction pointed out above)
\\[-1.7em]
\begin{align}
\small
\begin{squarecells}{3}
 ~~& $\lambda_0$ & $\lambda_1$ \nl
 $\lambda_0$ & \ka $\lambda_0$ & \kb $\lambda_1$ \nl
$\lambda_1$ & \kb $\lambda_1$ & \ka $\lambda_1\lambda_1$\nl
\end{squarecells}
\end{align}
only the last entry is left to be determined.\newpage

Since $\lambda_1\lambda_1 \in \bar{S}_0=\{0_S,\lambda_0\}$, there are clearly two choices for closing it
\begin{align}
\small
\begin{squarecells}{3}
 ISO& $\lambda_0$ & $\lambda_1$ \nl
 $\lambda_0$ & \ka $\lambda_0 $ & \kb $\lambda_1$ \nl
$\lambda_1$ & \kb $\lambda_1 $ & \ko $0_S$\nl
\end{squarecells}
\qquad
\begin{squarecells}{3}
 AdS& $\lambda_0$ & $\lambda_1$ \nl
 $\lambda_0$ & \ka $\lambda_0$ & \kb $\lambda_1$ \nl
$\lambda_1$ & \kb $\lambda_1$ & \ka $\lambda_0$\nl
\end{squarecells}
\end{align}
which effectively correspond to the Poincar\'{e} and AdS algebras, as the commutator
\begin{align}
[P_a,P_b]=\lambda_1\lambda_1[\tilde{P}_a,\tilde{P}_b]=\lambda_1\lambda_1\tilde{J}_{ab}
\end{align}
leads to $[P_a,P_b]=0$ and $[P_a,P_b]=J_{ab}$. However, this is not the only possibility!

\section{Maxwell algebras}
We notice that $\lambda_1\lambda_1$ multiplication could be closed by the another element $ \lambda_2 \in S_{0}$. Obviously this extends our table by the another row and column with the additional multiplications to be determined. One can show that $\lambda_0 \lambda_2=\lambda_0( \lambda_1\lambda_1)=(\lambda_0 \lambda_1)\lambda_1=\lambda_2$, thus
\\[-1.2em]
\begin{align}
\small
\begin{squarecells}{4}
    ~~&     $\lambda_0$ & $\lambda_1$     & $\lambda_2$     \nl
$\lambda_0$ & \ka $\lambda_0$ & \kb $\lambda_1$ & \ka $\lambda_2$ \nl
$\lambda_1$ & \kb $\lambda_1$ & \ka $\lambda_2$ & $~$       \nl
$\lambda_2$ & \ka $\lambda_2$ & $~$ $~$   & $~$       \nl 
\end{squarecells}
\qquad
\begin{squarecells}{4}
    ~~&     $\lambda_0$ & $\lambda_1$     & $\lambda_2$     \nl
$\lambda_0$ & \ka $\lambda_0$ & \kb $\lambda_1$ & \ka $\lambda_2$ \nl
$\lambda_1$ & \kb $\lambda_1$ & \ka $\lambda_2$ & \kb $\lambda_1 \lambda_2$ \nl
$\lambda_2$ & \ka $\lambda_2$ & \kb $\lambda_2 \lambda_1$ & \ka $\lambda_2\lambda_2$ \nl 
\end{squarecells}
\end{align}
Now it is easy to see that the unknown entries will be expressed by the powers of the $\lambda_1$ element. Because the resonant decomposition forces $\lambda_1^3\in \bar{S}_{1}$, we can immediately establish for the available list of elements the two possible associative tables for $\lambda_1^3=0_S$ and $\lambda_1^3=\lambda_1$ along with the straightforward identification of the corresponding $\lambda_1^4=0$ and  $\lambda_1^4=\lambda_2$
\\[-1.2em]
\begin{align}
\small
\begin{squarecells}{4}
    ~~&     $\lambda_0$ & $\lambda_1$       & $\lambda_2$     \nl
$\lambda_0$ & \ka $\lambda_0$ & \kb $\lambda_1$   & \ka $\lambda_2$ \nl
$\lambda_1$ & \kb $\lambda_1$ & \ka $\lambda_2$   & \kb $\lambda_1^3$ \nl
$\lambda_2$ & \ka $\lambda_2$ & \kb $\lambda_1^3$ & \ka $\lambda_1^4$ \nl 
\end{squarecells}
\qquad\qquad\qquad
\begin{squarecells}{4}
$\mathfrak{B}_4$  &     $\lambda_0$ & $\lambda_1$     & $\lambda_2$     \nl
$\lambda_0$ & \ka $\lambda_0$ & \kb $\lambda_1$ & \ka $\lambda_2$ \nl
$\lambda_1$ & \kb $\lambda_1$ & \ka $\lambda_2$ & \kb $0_S$ \nl
$\lambda_2$ & \ka $\lambda_2$ & \kb $0_S$ & \ka $0_S$ \nl 
\end{squarecells}
\qquad
\begin{squarecells}{4}
$\mathfrak{C}_4$  &     $\lambda_0$ & $\lambda_1$     & $\lambda_2$     \nl
$\lambda_0$ & \ka $\lambda_0$ & \kb $\lambda_1$ & \ka $\lambda_2$ \nl
$\lambda_1$ & \kb $\lambda_1$ & \ka $\lambda_2$ & \kb $\lambda_1$ \nl
$\lambda_2$ & \ka $\lambda_2$ & \kb $\lambda_1$ & \ka $\lambda_2$ \nl 
\end{squarecells}
\end{align}
Remarkably, these tables represent the Maxwell algebra of type  $\mathfrak{B}_4$  and $\mathfrak{C}_4\equiv AdS\oplus Lorentz$. In both cases the Lorentz $J_{ab}=J_{ab,(0)}=\lambda_0\tilde{J}_{ab}$ and translation $P_{a}=P_{a,(0)}=\lambda_1\tilde{P}_a$ generators are equipped with the new generator $Z_{ab}=J_{ab,(1)}=\lambda_2\tilde{J}_{ab}$ (see Appendix \ref{AppA} for more details). Besides starting papers \cite{Schrader:1972zd,Bacry:1970ye,Soroka:2004fj}, both algebras have been studied in various contexts, starting with the deformations and dynamical realizations
\cite{Gomis:2009dm}, use in the BF gravity models \cite{Durka:2011nf,Durka:2012wd} and exploiting the Maxwell symmetry in emergence of the cosmological constant term \cite{deAzcarraga:2010sw}. Further applications included a bimetric and cosmology context \cite{Durka:2011va} or construction of the Brans-Dicke theory \cite{Inostroza:2014vua}, WZW model \cite{Salgado:2014jka}, along with finding their supersymmetric extensions \cite{Soroka:2006aj,Durka:2011gm,Kamimura:2011mq,deAzcarraga:2012zv}.

Interestingly, the $\mathfrak{C}_4$ algebra under the specific change of basis can be rewritten as a direct sum of the AdS and Lorentz algebras (check Appendix \ref{AppB}). Such feature leads to the question if a commutator of any \textit{physical} Lorentz generator with any other generator could give zero. In this paper we decide to extend a behavior known from $[J,J]\sim \eta J$ and $[J,P]\sim \eta P$ in the further cases, leaving a discussion concerning other options for the future.

In analogy, the previously discussed multiplication of $\lambda_1 \lambda_2=\lambda_2 \lambda_1=\lambda_1^3$ could also lead to the new element $\lambda_3\in S_1$ associated with the new translational-like generator $R_{a}=P_{a,(1)}=\lambda_3\tilde{P}_a$ preserved by the Lorentz generator due to $\lambda_0 \lambda_3=(\lambda_0\lambda_1) \lambda_2=\lambda_3$. Obtained in this way table
\begin{align}
\small
\begin{squarecells}{5}
    ~~&     $\lambda_0$ & $\lambda_1$         & $\lambda_2$         & $\lambda_3$ \nl
$\lambda_0$ & \ka $\lambda_0$ & \kb $\lambda_1$     & \ka $\lambda_2$     & \kb $\lambda_3$ \nl
$\lambda_1$ & \kb $\lambda_1$ & \ka $\lambda_2$     & \kb $\lambda_3$     & \ka $\lambda_1 \lambda_3$ \nl
$\lambda_2$ & \ka $\lambda_2$ & \kb $\lambda_3$     & \ka $\lambda_2 \lambda_2$ & \kb $\lambda_2 \lambda_3$ \nl 
$\lambda_3$ & \kb $\lambda_3$ & \ka $\lambda_3 \lambda_1$ & \kb $\lambda_3 \lambda_2$ & \ka $\lambda_3 \lambda_3$ \nl
\end{squarecells}
\end{align}
due to the outcome of $\lambda_1 \lambda_3\in \{0_S,\lambda_0,\lambda_2\} $ immediately gives three patterns. To better emphasize the resulting structure let us stop coloring of the resonant decomposition and now color the different types of the semigroup elements
\begin{align}
\small
\begin{squarecells}{5}
$\mathfrak{B}_5$  &     $\lambda_0$ & $\lambda_1$     & $\lambda_2$    & $\lambda_3$ \nl
$\lambda_0$ & \kx $\lambda_0$ & \ky $\lambda_1$ & \kw $\lambda_2$ & \kv $\lambda_3$ \nl
$\lambda_1$ & \ky $\lambda_1$ & \kw $\lambda_2$ & \kv $\lambda_3$ & \ko $0_S$ \nl
$\lambda_2$ & \kw $\lambda_2$ & \kv $\lambda_3$ & \ko $0_S$ & \ko $0_S$ \nl 
$\lambda_3$ & \kv $\lambda_3$ & \ko $0_S$   & \ko $0_S$ & \ko $0_S$ \nl
\end{squarecells}
\qquad
\begin{squarecells}{5}
$\mathfrak{C}_5$  &     $\lambda_0$ & $\lambda_1$     & $\lambda_2$    & $\lambda_3$ \nl
$\lambda_0$ & \kx $\lambda_0$ & \ky $\lambda_1$ & \kw $\lambda_2$ & \kv $\lambda_3$ \nl
$\lambda_1$ & \ky $\lambda_1$ & \kw $\lambda_2$ & \kv $\lambda_3$ & \kx $\lambda_0$ \nl
$\lambda_2$ & \kw $\lambda_2$ & \kv $\lambda_3$ & \kx $\lambda_0$ & \ky $\lambda_1$ \nl 
$\lambda_3$ & \kv $\lambda_3$ & \kx $\lambda_0$ & \ky $\lambda_1$ & \kw $\lambda_2$ \nl
\end{squarecells}
\qquad
\begin{squarecells}{5}
$\mathfrak{D}_5$  &     $\lambda_0$ & $\lambda_1$     & $\lambda_2$     & $\lambda_3$ \nl
$\lambda_0$ & \kx $\lambda_0$ & \ky $\lambda_1$ & \kw $\lambda_2$ & \kv $\lambda_3$ \nl
$\lambda_1$ & \ky $\lambda_1$ & \kw $\lambda_2$ & \kv $\lambda_3$ & \kw $\lambda_2$ \nl
$\lambda_2$ & \kw $\lambda_2$ & \kv $\lambda_3$ & \kw $\lambda_2$ & \kv $\lambda_3$ \nl 
$\lambda_3$ & \kv $\lambda_3$ & \kw $\lambda_2$ & \kv $\lambda_3$ & \kw $\lambda_2$ \nl
\end{squarecells}
\label{B5C5D5}
\end{align}
These three examples share the same part up to the main anti-diagonal (with the simple multiplication law $\lambda_\alpha \lambda_\beta=\lambda_{\alpha+\beta}$), whereas further we are going to have:
\begin{center}
\qquad\quad all zeros,
\qquad\qquad\qquad anti-diagonal pattern,
\qquad\qquad chessboard pattern.
\end{center}
In this way we have found semigroups corresponding to the Maxwell algebras of type $\mathfrak{B}$,~$\mathfrak{C}$,~$\mathfrak{D}$, which are spanned by four kinds of the generators $\{J_{ab},P_{a},Z_{ab},R_{a}\}$. The associated field content $\{\omega^{ab},\frac{1}{\ell}e^ {a},k^{ab},\frac{1}{\ell}h^ {a}\}$, through the expanded curvatures and modifications of the invariant tensors, affects non-trivially the final form of the Lagrangians. This was exploited in  Refs.~\cite{Edelstein:2006se,Izaurieta:2009hz} to establish though $\mathfrak{B}_5$ algebra a relation in $5D$ between Chern-Simons (CS) gravity and the Einstein-Hilbert action. After generalization to the arbitrary odd dimensions, the same was achieved in even dimensions for Born-Infeld (BI) gravity and GR \cite{Concha:2013uhq,Concha:2014vka}. Algebra $\mathfrak{C}_5$ was recently analyzed in a similar context to obtain the so-called Pure Lovelock (PL) action \cite{Concha:2016kdz}, which instead of the full Lanczos-Lovelock series contains only the cosmological constant term and a single $p$ power polynomial term in the Riemann curvature (in $5D$ it can be either $RRe$ or $Re^3$). The last $\mathfrak{D}_5$ algebra introduced in Ref.~\cite{Concha:2016hbt} admits the direct sum of the $AdS\oplus \textit{Poincar\'{e}}$ algebras (see Appendix~\ref{AppB}), which effectively leads to the CS/BI action expressed as the sum of two independent pieces.

Ultimately, we arrive at the arbitrary number of elements/generators. For algebra with $m=6$ one fixes $\lambda_1\lambda_3=\lambda^4_1$ as a new element, thus considering $\lambda_1^5 \in \{0_S,\lambda_1,\lambda_3\}$ produces tables:
\\[-1.5em]
\begin{align*}
\footnotesize
\begin{squarecells}{6}
$\mathfrak{B}_6$  &     $\lambda_0$ & $\lambda_1$     & $\lambda_2$    & $\lambda_3$ & $\lambda_4$ \nl
$\lambda_0$ & \kx $\lambda_0$ & \ky $\lambda_1$ & \kw $\lambda_2$ & \kv $\lambda_3$ & \ku $\lambda_4$ \nl
$\lambda_1$ & \ky $\lambda_1$ & \kw $\lambda_2$ & \kv $\lambda_3$ & \ku $\lambda_4$ & \ko $0_S$   \nl
$\lambda_2$ & \kw $\lambda_2$ & \kv $\lambda_3$ & \ku $\lambda_4$ & \ko $0_S$   & \ko $0_S$   \nl 
$\lambda_3$ & \kv $\lambda_3$ & \ku $\lambda_4$ & \ko $0_S$   & \ko $0_S$   & \ko $0_S$   \nl
$\lambda_4$ & \ku $\lambda_4$ & \ko $0_S$   & \ko $0_S$   & \ko $0_S$   & \ko $0_S$   \nl
\end{squarecells}
\quad
\begin{squarecells}{6}
$\mathfrak{C}_6$  &     $\lambda_0$ & $\lambda_1$     & $\lambda_2$    & $\lambda_3$ & $\lambda_4$ \nl
$\lambda_0$ & \kx $\lambda_0$ & \ky $\lambda_1$ & \kw $\lambda_2$ & \kv $\lambda_3$ & \ku $\lambda_4$ \nl
$\lambda_1$ & \ky $\lambda_1$ & \kw $\lambda_2$ & \kv $\lambda_3$ & \ku $\lambda_4$ & \ky $\lambda_1$ \nl
$\lambda_2$ & \kw $\lambda_2$ & \kv $\lambda_3$ & \ku $\lambda_4$ & \ky $\lambda_1$ & \kw $\lambda_2$ \nl 
$\lambda_3$ & \kv $\lambda_3$ & \ku $\lambda_4$ & \ky $\lambda_1$ & \kw $\lambda_2$ & \kv $\lambda_3$ \nl
$\lambda_4$ & \ku $\lambda_4$ & \ky $\lambda_1$ & \kw $\lambda_2$ & \kv $\lambda_3$ & \ku $\lambda_4$ \nl
\end{squarecells}
\quad
\begin{squarecells}{6}
$\mathfrak{D}_6$ &     $\lambda_0$ & $\lambda_1$     & $\lambda_2$    & $\lambda_3$ & $\lambda_4$ \nl
$\lambda_0$ & \kx $\lambda_0$ & \ky $\lambda_1$ & \kw $\lambda_2$ & \kv $\lambda_3$ & \ku $\lambda_4$ \nl
$\lambda_1$ & \ky $\lambda_1$ & \kw $\lambda_2$ & \kv $\lambda_3$ & \ku $\lambda_4$ & \kv $\lambda_3$ \nl
$\lambda_2$ & \kw $\lambda_2$ & \kv $\lambda_3$ & \ku $\lambda_4$ & \kv $\lambda_3$ & \ku $\lambda_4$ \nl 
$\lambda_3$ & \kv $\lambda_3$ & \ku $\lambda_4$ & \kv $\lambda_3$ & \ku $\lambda_4$ & \kv $\lambda_3$ \nl
$\lambda_4$ & \ku $\lambda_4$ & \kv $\lambda_3$ & \ku $\lambda_4$ & \kv $\lambda_3$ & \ku $\lambda_4$ \nl
\end{squarecells}
\end{align*}
For a larger number of elements it will be possible to find more schemes. Indeed, starting from $m=7$ we can find another family $\mathfrak{E}_m$, which structurally resembles a lot of the $\mathfrak{C}_m$ algebra:
\begin{align}
\footnotesize 
\begin{squarecells}{7}
$\mathfrak{B}_7$  &  $\lambda_0$    &  $\lambda_1$    & $\lambda_2$     & $\lambda_3$     & $\lambda_4$ &  $\lambda_5$ \nl
$\lambda_0$ & \kx $\lambda_0$ & \ky $\lambda_1$ & \kw $\lambda_2$ & \kv $\lambda_3$ & \ku $\lambda_4$ & \kq $\lambda_5$ \nl
$\lambda_1$ & \ky $\lambda_1$ & \kw $\lambda_2$ & \kv $\lambda_3$ & \ku $\lambda_4$ & \kq $\lambda_5$ & \ko $0_S$ \nl
$\lambda_2$ & \kw $\lambda_2$ & \kv $\lambda_3$ & \ku $\lambda_4$ & \kq $\lambda_5$ & \ko $0_S$   & \ko $0_S$ \nl 
$\lambda_3$ & \kv $\lambda_3$ & \ku $\lambda_4$ & \kq $\lambda_5$ & \ko $0_S$   & \ko $0_S$   & \ko $0_S$ \nl
$\lambda_4$ & \ku $\lambda_4$ & \kq $\lambda_5$ & \ko $0_S$   & \ko $0_S$   & \ko $0_S$   & \ko $0_S$ \nl
$\lambda_5$ & \kq $\lambda_5$ & \ko $0_S$   & \ko $0_S$   & \ko $0_S$   & \ko $0_S$   & \ko $0_S$ \nl
\end{squarecells}
\qquad
\begin{squarecells}{7}
$\mathfrak{C}_7$  &  $\lambda_0$    &  $\lambda_1$    & $\lambda_2$     & $\lambda_3$     & $\lambda_4$ &  $\lambda_5$ \nl
$\lambda_0$ & \kx $\lambda_0$ & \ky $\lambda_1$ & \kw $\lambda_2$ & \kv $\lambda_3$ & \ku $\lambda_4$ & \kq $\lambda_5$ \nl
$\lambda_1$ & \ky $\lambda_1$ & \kw $\lambda_2$ & \kv $\lambda_3$ & \ku $\lambda_4$ & \kq $\lambda_5$ & \kx $\lambda_0$ \nl
$\lambda_2$ & \kw $\lambda_2$ & \kv $\lambda_3$ & \ku $\lambda_4$ & \kq $\lambda_5$ & \kx $\lambda_0$ & \ky $\lambda_1$ \nl
$\lambda_3$ & \kv $\lambda_3$ & \ku $\lambda_4$ & \kq $\lambda_5$ & \kx $\lambda_0$ & \ky $\lambda_1$ & \kw $\lambda_2$ \nl
$\lambda_4$ & \ku $\lambda_4$ & \kq $\lambda_5$ & \kx $\lambda_0$ & \ky $\lambda_1$ & \kw $\lambda_2$ & \kv $\lambda_3$ \nl
$\lambda_5$ & \kq $\lambda_5$ & \kx $\lambda_0$ & \ky $\lambda_1$ & \kw $\lambda_2$ & \kv $\lambda_3$ & \ku $\lambda_4$ \nl
\end{squarecells}\nonumber\\
\footnotesize
\begin{squarecells}{7}
$\mathfrak{D}_7$ &  $\lambda_0$    &  $\lambda_1$    & $\lambda_2$     & $\lambda_3$     & $\lambda_4$ &  $\lambda_5$ \nl
$\lambda_0$ & \kx $\lambda_0$ & \ky $\lambda_1$ & \kw $\lambda_2$ & \kv $\lambda_3$ & \ku $\lambda_4$ & \kq $\lambda_5$ \nl
$\lambda_1$ & \ky $\lambda_1$ & \kw $\lambda_2$ & \kv $\lambda_3$ & \ku $\lambda_4$ & \kq $\lambda_5$ & \ku $\lambda_4$ \nl
$\lambda_2$ & \kw $\lambda_2$ & \kv $\lambda_3$ & \ku $\lambda_4$ & \kq $\lambda_5$ & \ku $\lambda_4$ & \kq $\lambda_5$ \nl
$\lambda_3$ & \kv $\lambda_3$ & \ku $\lambda_4$ & \kq $\lambda_5$ & \ku $\lambda_4$ & \kq $\lambda_5$ & \ku $\lambda_4$ \nl
$\lambda_4$ & \ku $\lambda_4$ & \kq $\lambda_5$ & \ku $\lambda_4$ & \kq $\lambda_5$ & \ku $\lambda_4$ & \kq $\lambda_5$ \nl
$\lambda_5$ & \kq $\lambda_5$ & \ku $\lambda_4$ & \kq $\lambda_5$ & \ku $\lambda_4$ & \kq $\lambda_5$ & \ku $\lambda_4$ \nl
\end{squarecells}
\qquad
\begin{squarecells}{7}
$\mathfrak{E}_7$   &  $\lambda_0$    &  $\lambda_1$    & $\lambda_2$     & $\lambda_3$     & $\lambda_4$ &  $\lambda_5$ \nl
$\lambda_0$ & \kx $\lambda_0$ & \ky $\lambda_1$ & \kw $\lambda_2$ & \kv $\lambda_3$ & \ku $\lambda_4$ & \kq $\lambda_5$ \nl
$\lambda_1$ & \ky $\lambda_1$ & \kw $\lambda_2$ & \kv $\lambda_3$ & \ku $\lambda_4$ & \kq $\lambda_5$ & \kw $\lambda_2$ \nl
$\lambda_2$ & \kw $\lambda_2$ & \kv $\lambda_3$ & \ku $\lambda_4$ & \kq $\lambda_5$ & \kw $\lambda_2$ & \kv $\lambda_3$ \nl
$\lambda_3$ & \kv $\lambda_3$ & \ku $\lambda_4$ & \kq $\lambda_5$ & \kw $\lambda_2$ & \kv $\lambda_3$ & \ku $\lambda_4$ \nl
$\lambda_4$ & \ku $\lambda_4$ & \kq $\lambda_5$ & \kw $\lambda_2$ & \kv $\lambda_3$ & \ku $\lambda_4$ & \kq $\lambda_5$ \nl
$\lambda_5$ & \kq $\lambda_5$ & \kw $\lambda_2$ & \kv $\lambda_3$ & \ku $\lambda_4$ & \kq $\lambda_5$ & \kw $\lambda_2$ \nl
\end{squarecells}\label{BCDE7}
\end{align}
The total number of possible patterns depends on $m$ and is equal to the integer part of $\left[ \frac{m+1}{2}\right]$, which comes from fixing the outcome of $\lambda_1^{m-1}$ as being one of $\{0_S,\lambda_0,\lambda_2,...,\lambda_{m-3}\}$ when $m=odd$ or  $\{0_S,\lambda_1,\lambda_3,...,\lambda_{m-3}\}$ when $m=even$. Excluding the $\mathfrak{B}_m$ family (the scheme with $0_S$), it is possible to capture all other Maxwell families in the single multiplication law
\begin{equation}
\lambda_{\alpha}\lambda_{\beta}=\left\{
\begin{array}
[c]{l}%
\lambda_{\alpha+\beta}\,,\qquad~~~
\text{for\quad}\alpha+\beta\leq m-2\\
\lambda_\gamma\,,
\qquad\qquad\,\text{for\quad}\alpha+\beta>m-2
\end{array}
\right.
\end{equation}
where element $\lambda_\rho$, being directly under the main anti-diagonal, determines the particular type of the algebra through
\begin{equation}
\gamma=
\left(\alpha+\beta-(m-1)\right)\operatorname{mod}\left((m-1)-\rho\right)+\rho\,.
\end{equation}
Family $\mathfrak{D}_{m\geq 5}$ with the maximal element indicated by $\rho=m-3$ reproduces the chessboard scheme, where the rest of the families realize an anti-diagonal arrangement differing only in the starting element. Regarding the parity and not exceeding the range fixed by the existence of the $\mathfrak{D}_{m}$, the algebra of type $\mathfrak{C}_{m\geq 3}$ will be generated by $\rho=0$ ($m=odd$) and $\rho=1$ ($m=even$). Analogously, if $m$ is large enough, $\rho=2$ and $\rho=3$ will generate $\mathfrak{E}_{m\geq 7}$, whereas $\rho=4$ and $\rho=5$ will generate $\mathfrak{F}_{m\geq 9}$, and so on. Finally, the $\mathfrak{B}_m$ family is retrieved as the In\"{o}n\"{u}-Wigner contraction of all these algebras in the limit of dimensionless parameter $\mu\rightarrow\infty$ scaling generators $P_{a,(0)}\rightarrow \mu P_{a,(0)}\,,J_{ab,\left(  1\right)}\rightarrow \mu^2 J_{ab,\left(  1\right)}\,,P_{a,(1)}\rightarrow \mu^3 P_{a,(1)},...$

\section{Poincar\'{e}-like, AdS-like, and Maxwell-like algebras}
It is important to note that in our derivation we have missed quite an interesting branch of enlargements coming from the possibility of adding the new element $\lambda_2$ independently to the table, while still having $\lambda_1 \lambda_1$ equal to $0_S$ or $\lambda_0$. This goes beyond the procedure presented earlier, as the new elements will not be related to the powers of $\lambda_1$. Then, to obtain the proper tables there is nothing else but to laboriously analyze the semigroup associativity. Considering the general setup, we see the emergence of the three basic families containing the Poincar\'{e}, the AdS and the Maxwell algebras as the subalgebra:
\\[-1.6em]
\begin{align}
\small
\begin{squarecells}{4}
    ~~&     $\lambda_0$ & $\lambda_1$     & $\lambda_2$     \nl
$\lambda_0$ & \ka $\lambda_0$ & \kb $\lambda_1$ & \ka $\lambda_2$ \nl
$\lambda_1$ & \kb $\lambda_1$ & \ks $0_S$ & \kb $~~$ \nl
$\lambda_2$ & \ka $\lambda_2$ & \kb $~~$ & \ka $~~$ \nl 
\end{squarecells}
\qquad
\begin{squarecells}{4}
    ~~&     $\lambda_0$ & $\lambda_1$     & $\lambda_2$     \nl
$\lambda_0$ & \ka $\lambda_0$ & \kb $\lambda_1$ & \ka $\lambda_2$ \nl
$\lambda_1$ & \kb $\lambda_1$ & \ks $\lambda_0$ & \kb $~~$ \nl
$\lambda_2$ & \ka $\lambda_2$ & \kb $~~$ & \ka $~~$ \nl 
\end{squarecells}
\qquad
\begin{squarecells}{4}
    ~~&     $\lambda_0$ & $\lambda_1$     & $\lambda_2$     \nl
$\lambda_0$ & \ka $\lambda_0$ & \kb $\lambda_1$ & \ka $\lambda_2$ \nl
$\lambda_1$ & \kb $\lambda_1$ & \ks $\lambda_2$ & \kb $~~$ \nl
$\lambda_2$ & \ka $\lambda_2$ & \kb $~~$ & \ka $~~$ \nl 
\end{squarecells}
\end{align}
Of course this generalization could go on further, i.e. $\lambda_1 \lambda_1=\lambda_4$, then $\lambda_1 \lambda_1=\lambda_6$, and so on, but we will not explore such scenarios here.\\

\noindent An explicit check of the associativity (see dedicated tool at webpage \textit{resonantalgebras.wordpress.com}) shows that it is possible to construct:
\begin{itemize}
\item $4 \times$ Poincar\'{e}-like algebras we could denote as type $B_4$, $\tilde{B}_4$, $\tilde{C}_4$, and $C_4 \equiv ISO\oplus Lorentz$:
\\[-1.6em]
\begin{align}
\small
\begin{squarecells}{4}
$B_4$ &     $\lambda_0$ & $\lambda_1$     & $\lambda_2$     \nl
$\lambda_0$ & \kx $\lambda_0$ & \ky $\lambda_1$ & \kw $\lambda_2$ \nl
$\lambda_1$ & \ky $\lambda_1$ & \ko $0_S$ & \ko $0_S$ \nl
$\lambda_2$ & \kw $\lambda_2$ & \ko $0_S$ & \ko $0_S$ \nl 
\end{squarecells}
\,
\begin{squarecells}{4}
$\tilde{B}_4$ &      $\lambda_0$ & $\lambda_1$     & $\lambda_2$     \nl
$\lambda_0$ & \kx $\lambda_0$ & \ky $\lambda_1$ & \kw $\lambda_2$ \nl
$\lambda_1$ & \ky $\lambda_1$ & \ko $0_S$ & \ko $0_S$ \nl
$\lambda_2$ & \kw $\lambda_2$ & \ko $0_S$ & \kw $\lambda_2$ \nl 
\end{squarecells}
\,
\begin{squarecells}{4}
$\tilde{C}_4$ &      $\lambda_0$ & $\lambda_1$     & $\lambda_2$     \nl
$\lambda_0$ & \kx $\lambda_0$ & \ky $\lambda_1$ & \kw $\lambda_2$ \nl
$\lambda_1$ & \ky $\lambda_1$ & \ko $0_S$ & \ky $\lambda_1$ \nl
$\lambda_2$ & \kw $\lambda_2$ & \ky $\lambda_1$ & \ko $0_S$ \nl 
\end{squarecells}
\,
\begin{squarecells}{4}
$C_4$ &      $\lambda_0$ & $\lambda_1$     & $\lambda_2$     \nl
$\lambda_0$ & \kx $\lambda_0$ & \ky $\lambda_1$ & \kw $\lambda_2$ \nl
$\lambda_1$ & \ky $\lambda_1$ & \ko $0_S$ & \ky $\lambda_1$ \nl
$\lambda_2$ & \kw $\lambda_2$ & \ky $\lambda_1$ & \kw $\lambda_2$ \nl 
\end{squarecells}
\label{Poincare-like-m4}
\end{align}
\item none of the AdS-like algebra (the associativity is not fulfilled in any configuration)
\item $2 \times$ Maxwell-like algebras of type $\mathfrak{B}_4$ and $\mathfrak{C}_4$ already introduced in the previous section:
\begin{align}
\small
\begin{squarecells}{4}
$\mathfrak{B}_4$  &     $\lambda_0$ & $\lambda_1$     & $\lambda_2$     \nl
$\lambda_0$ & \kx $\lambda_0$ & \ky $\lambda_1$ & \kw $\lambda_2$ \nl
$\lambda_1$ & \ky $\lambda_1$ & \kw $\lambda_2$ & \ko $0_S$ \nl
$\lambda_2$ & \kw $\lambda_2$ & \ko $0_S$ & \ko $0_S$ \nl 
\end{squarecells}
\qquad
\begin{squarecells}{4}
$\mathfrak{C}_4$  &     $\lambda_0$ & $\lambda_1$     & $\lambda_2$     \nl
$\lambda_0$ & \kx $\lambda_0$ & \ky $\lambda_1$ & \kw $\lambda_2$ \nl
$\lambda_1$ & \ky $\lambda_1$ & \kw $\lambda_2$ & \ky $\lambda_1$ \nl
$\lambda_2$ & \kw $\lambda_2$ & \ky $\lambda_1$ & \kw $\lambda_2$ \nl 
\end{squarecells}
\end{align}
   \end{itemize}
Further enlargement, coming with the new translational $R_a=P_{a,(1)}=\lambda_3 \tilde{P}_a$ generator, brings a much richer collection of the algebras:
\begin{itemize}
\item $17 \times$ Poincar\'{e}-like
\item $3 \times$  AdS-like of type $\mathcal{B}_5$, $\mathcal{C}_5$, and $\mathcal{D}_5\equiv AdS\oplus AdS$
\\[-1.2em]
\begin{align}
\small
\begin{squarecells}{5}
$\mathcal{B}_5$  &     $\lambda_0$ & $\lambda_1$     & $\lambda_2$    & $\lambda_3$ \nl
$\lambda_0$ & \kx $\lambda_0$ & \ky $\lambda_1$ & \kw $\lambda_2$ & \kv $\lambda_3$ \nl
$\lambda_1$ & \ky $\lambda_1$ & \kx $\lambda_0$ & \kv $\lambda_3$ & \kw $\lambda_2$ \nl
$\lambda_2$ & \kw $\lambda_2$ & \kv $\lambda_3$ & \ko $0_S$ & \ko $0_S$ \nl 
$\lambda_3$ & \kv $\lambda_3$ & \kw $\lambda_2$   & \ko $0_S$ & \ko $0_S$ \nl
\end{squarecells}
\quad
\begin{squarecells}{5}
 $\mathcal{C}_5$&     $\lambda_0$ & $\lambda_1$     & $\lambda_2$    & $\lambda_3$ \nl
$\lambda_0$ & \kx $\lambda_0$ & \ky $\lambda_1$ & \kw $\lambda_2$ & \kv $\lambda_3$ \nl
$\lambda_1$ & \ky $\lambda_1$ & \kx $\lambda_0$ & \kv $\lambda_3$ & \kw $\lambda_2$ \nl
$\lambda_2$ & \kw $\lambda_2$ & \kv $\lambda_3$ & \kx $\lambda_0$ & \ky $\lambda_1$ \nl 
$\lambda_3$ & \kv $\lambda_3$ & \kw $\lambda_2$ & \ky $\lambda_1$ & \kx $\lambda_0$ \nl
\end{squarecells}
\quad
\begin{squarecells}{5}
$\mathcal{D}_5$  &     $\lambda_0$ & $\lambda_1$     & $\lambda_2$    & $\lambda_3$ \nl
$\lambda_0$ & \kx $\lambda_0$ & \ky $\lambda_1$ & \kw $\lambda_2$ & \kv $\lambda_3$ \nl
$\lambda_1$ & \ky $\lambda_1$ & \kx $\lambda_0$ & \kv $\lambda_3$ & \kw $\lambda_2$ \nl
$\lambda_2$ & \kw $\lambda_2$ & \kv $\lambda_3$ & \kw $\lambda_2$ & \kv $\lambda_3$ \nl 
$\lambda_3$ & \kv $\lambda_3$ & \kw $\lambda_2$ & \kv $\lambda_3$ & \kw $\lambda_2$ \nl
\end{squarecells}\label{AdS-like-1}
\end{align}
\item $10 \times$ Maxwell-like: three of them $\mathfrak{B}_5,\mathfrak{C}_5,\mathfrak{D}_5$ were already derived in a previous section
\begin{align}
\small
\begin{squarecells}{5}
$\mathfrak{B}_5$  &     $\lambda_0$ & $\lambda_1$     & $\lambda_2$    & $\lambda_3$ \nl
$\lambda_0$ & \kx $\lambda_0$ & \ky $\lambda_1$ & \kw $\lambda_2$ & \kv $\lambda_3$ \nl
$\lambda_1$ & \ky $\lambda_1$ & \kw $\lambda_2$ & \kv $\lambda_3$ & \ko $0_S$ \nl
$\lambda_2$ & \kw $\lambda_2$ & \kv $\lambda_3$ & \ko $0_S$ & \ko $0_S$ \nl 
$\lambda_3$ & \kv $\lambda_3$ & \ko $0_S$ & \ko $0_S$ & \ko $0_S$ \nl
\end{squarecells}
\quad
\begin{squarecells}{5}
$\mathfrak{C}_5$  &     $\lambda_0$ & $\lambda_1$     & $\lambda_2$    & $\lambda_3$ \nl
$\lambda_0$ & \kx $\lambda_0$ & \ky $\lambda_1$ & \kw $\lambda_2$ & \kv $\lambda_3$ \nl
$\lambda_1$ & \ky $\lambda_1$ & \kw $\lambda_2$ & \kv $\lambda_3$ & \kx $\lambda_0$ \nl
$\lambda_2$ & \kw $\lambda_2$ & \kv $\lambda_3$ & \kx $\lambda_0$ & \ky $\lambda_1$ \nl 
$\lambda_3$ & \kv $\lambda_3$ & \kx $\lambda_0$ & \ky $\lambda_1$ & \kw $\lambda_2$ \nl
\end{squarecells}
\quad
\begin{squarecells}{5}
$\mathfrak{D}_5$  &     $\lambda_0$ & $\lambda_1$     & $\lambda_2$    & $\lambda_3$ \nl
$\lambda_0$ & \kx $\lambda_0$ & \ky $\lambda_1$ & \kw $\lambda_2$ & \kv $\lambda_3$ \nl
$\lambda_1$ & \ky $\lambda_1$ & \kw $\lambda_2$ & \kv $\lambda_3$ & \kw $\lambda_2$ \nl
$\lambda_2$ & \kw $\lambda_2$ & \kv $\lambda_3$ & \kw $\lambda_2$ & \kv $\lambda_3$ \nl 
$\lambda_3$ & \kv $\lambda_3$ & \kw $\lambda_2$ & \kv $\lambda_3$ & \kw $\lambda_2$ \nl
\end{squarecells}
\end{align}
but surprisingly there are five others with the $0_S$ (from them only one is being presented below) and an additional two without the zero elements
\begin{align}
\small
\begin{squarecells}{5}
$~~$  &     $\lambda_0$ & $\lambda_1$     & $\lambda_2$    & $\lambda_3$ \nl
$\lambda_0$ & \kx $\lambda_0$ & \ky $\lambda_1$ & \kw $\lambda_2$ & \kv $\lambda_3$ \nl
$\lambda_1$ & \ky $\lambda_1$ & \kw $\lambda_2$ & \ky $\lambda_1$ & \ko $0_S$ \nl
$\lambda_2$ & \kw $\lambda_2$ & \ky $\lambda_1$ & \kw $\lambda_2$ & \ko $0_S$ \nl 
$\lambda_3$ & \kv $\lambda_3$ & \ko $0_S$ & \ko $0_S$ & \ko $0_S$ \nl
\end{squarecells}
\quad
\begin{squarecells}{5}
$~~$  &     $\lambda_0$ & $\lambda_1$     & $\lambda_2$    & $\lambda_3$ \nl
$\lambda_0$ & \kx $\lambda_0$ & \ky $\lambda_1$ & \kw $\lambda_2$ & \kv $\lambda_3$ \nl
$\lambda_1$ & \ky $\lambda_1$ & \kw $\lambda_2$ & \ky $\lambda_1$ & \kw $\lambda_2$ \nl
$\lambda_2$ & \kw $\lambda_2$ & \ky $\lambda_1$ & \kw $\lambda_2$ & \ky $\lambda_1$ \nl 
$\lambda_3$ & \kv $\lambda_3$ & \kw $\lambda_2$ & \ky $\lambda_1$ & \kw $\lambda_2$ \nl
\end{squarecells}
\quad
\begin{squarecells}{5}
$~~$  &     $\lambda_0$ & $\lambda_1$     & $\lambda_2$    & $\lambda_3$ \nl
$\lambda_0$ & \kx $\lambda_0$ & \ky $\lambda_1$ & \kw $\lambda_2$ & \kv $\lambda_3$ \nl
$\lambda_1$ & \ky $\lambda_1$ & \kw $\lambda_2$ & \ky $\lambda_1$ & \kw $\lambda_2$ \nl
$\lambda_2$ & \kw $\lambda_2$ & \ky $\lambda_1$ & \kw $\lambda_2$ & \ky $\lambda_1$ \nl 
$\lambda_3$ & \kv $\lambda_3$ & \kw $\lambda_2$ & \ky $\lambda_1$ & \kx $\lambda_0$ \nl
\end{squarecells}\label{Max-like-2}
\end{align}
\end{itemize}
Normal, calligraphic and gothic labels distinguish, respectively, Poincar\'{e}, AdS, and Maxwell like algebras. Note that the elements of the semigroup $\mathfrak{C}_5$, just like any other representative of this family when $m=odd$, correspond to the cyclic group $\mathbb{Z}_{4}$ (in arbitrary case $\mathbb{Z}_{m-1}$). In turn, one of the AdS-like tables, schematically denoted as $\mathcal{C}_5$, corresponds to the Klein group given by $\mathbb{Z}_{2}\times \mathbb{Z}_{2}$. It appeared in the $S$-expansion context in Ref.~\cite{Gonzalez:2014tta}, but some special conditions introducing minus signs to the definitions of the generators were necessary to reach its final goal. 

Explicit tables for the algebras labeled by $m=3,4,5,6$, with or without the zero elements, as well as the tool checking the
associativity, can be found at \textit{resonantalgebras.wordpress.com}. 

Before going any further, it should be pointed out that in gravity applications, the spin connection should be associated with the semigroup identity (Lorentz generator), but there might still be an ambiguity related to what is the true vielbein. Switching the labels between $e^a$ and $h^a$ and interchanging the second row and column with the forth row and column causes an isomorphism between the last AdS-like table in (\ref{AdS-like-1}) and the last Maxwell-like example in (\ref{Max-like-2}), as well as between the first AdS-like and some other Poincar\'{e}-like table. Moreover, the same could be said about $4 \times$Poincar\'{e}-like tables and $4\times$Maxwell-like containing the zero elements. To avoid confusion while constructing the gravity theory and not to complicate the classification of the algebras, we are going to always relate a vielbein $e^a$ to $\lambda_1$ and $P_a$, just like spin connection $\omega^{ab}$ will be always related to $\lambda_0$ and the Lorentz generator $J_{ab}$.

\section{Non-standard enlargement}

It is worth to mention one more overlooked generalization, which lies in the enlargement order. In principle there could be two schemes for adding the new generators:
\begin{itemize}
\item $J_{ab},P_{a}$ and then $Z_{ab}$
\item $J_{ab},P_{a}$ and then $R_{a}$
\end{itemize}
but only first one was explored in the literature. Of course including the additional generator $R_{a}$ to the first scheme as in a standard enlargement and adding $Z_{ab}$ to the second gives the same result, as $\{J_{ab},P_{a},R_{a},Z_{ab}\}\equiv \{J_{ab},P_{a},Z_{ab},R_{a}\}$. However, for yet another generator we return to the same starting situation. Minimal setup with only three non-zero elements $\{\lambda_0, \lambda_1,\lambda_3\}$ shows that semigroup conditions used in this paper are fulfilled just in one case given below under the label \textit{I}. Loosening up for a moment the requirement of an identity element, so $\lambda_0$ does
not necessarily preserves anything other than $\lambda_0$ and $\lambda_1$, brings two more tables:
\begin{align*}
\tiny
\begin{squarecells}{4}
  I  ~~&     $\lambda_0$ & $\lambda_1$     & $\lambda_3$     \nl
$\lambda_0$ & \ka $\lambda_0$ & \kb $\lambda_1$ & \kb $\lambda_3$ \nl
$\lambda_1$ & \kb $\lambda_1$ & \ka $0_S$ & \ka $0_S$ \nl
$\lambda_3$ & \kb $\lambda_3$ & \ka $0_S$ & \ka $0_S$ \nl 
\end{squarecells}
\qquad
\begin{squarecells}{4}
  II ~~&     $\lambda_0$ & $\lambda_1$     & $\lambda_3$     \nl
$\lambda_0$ & \ka $\lambda_0$ & \kb $\lambda_1$ & \kb $\lambda_1$ \nl
$\lambda_1$ & \kb $\lambda_1$ & \ka $0_S$ & \ka $0_S$ \nl
$\lambda_3$ & \kb $\lambda_1$ & \ka $0_S$ & \ka $0_S$ \nl 
\end{squarecells}
\qquad
\begin{squarecells}{4}
 III   ~~&     $\lambda_0$ & $\lambda_1$     & $\lambda_3$     \nl
$\lambda_0$ & \ka $\lambda_0$ & \kb $\lambda_1$ & \kb $\lambda_1$ \nl
$\lambda_1$ & \kb $\lambda_1$ & \ka $\lambda_0$ & \ka $\lambda_0$ \nl
$\lambda_3$ & \kb $\lambda_1$ & \ka $\lambda_0$ & \ka $\lambda_0$ \nl 
\end{squarecells}
\end{align*}
Related curvatures are defined as:
\small
\begin{align}
F_I&=\frac{1}{2}R^{ab}J_{ab}+\frac{1}{\ell}(de^a +\omega^{a}_{~b} e^b)P_a+\frac{1}{\ell}( dh^a +\omega^{a}_{~b} h^b)R_a\,,\\
F_{II}&=\frac{1}{2}R^{ab}J_{ab}+\frac{1}{\ell}( de^a +\omega^{a}_{~b} e^b+\omega^{a}_{~b} h^b)P_a+\frac{1}{\ell} dh^a R_a\,,\\
F_{III}&=\frac{1}{2}\left(R^{ab}+\frac{1}{\ell^2}(e^a+h^a) (e^b+h^b)\right)J_{ab}+\frac{1}{\ell}( de^a +\omega^{a}_{~b} e^b+\omega^{a}_{~b} h^b)P_a+\frac{1}{\ell} dh^a R_a\,.
\end{align}
\normalsize
The last case opens an interesting possibility of the extra contribution to a cosmological constant with the price to pay in a modification of the torsion and possible impact on the boundary. It would be very interesting to understand the implications of giving away also a preservation of $\lambda_0\lambda_1$ outcome, which naturally would have even more serious consequences for the translations and the definition of torsion. Such associative tables read
\begin{align*}
\tiny
\begin{squarecells}{4}
    ~~&     $\lambda_0$ & $\lambda_1$     & $\lambda_3$     \nl
$\lambda_0$ & \ka $\lambda_0$ & \kb $\lambda_3$ & \kb $\lambda_3$ \nl
$\lambda_1$ & \kb $\lambda_3$ & \ka $0_S$ & \ka $0_S$ \nl
$\lambda_3$ & \kb $\lambda_3$ & \ka $0_S$ & \ka $0_S$ \nl 
\end{squarecells}
\qquad
\begin{squarecells}{4}
    ~~&     $\lambda_0$ & $\lambda_1$     & $\lambda_3$     \nl
$\lambda_0$ & \ka $\lambda_0$ & \kb $\lambda_3$ & \kb $\lambda_3$ \nl
$\lambda_1$ & \kb $\lambda_3$ & \ka $\lambda_0$ & \ka $\lambda_0$ \nl
$\lambda_3$ & \kb $\lambda_3$ & \ka $\lambda_0$ & \ka $\lambda_0$ \nl 
\end{squarecells}
\end{align*}
causing effectively reverting the labels of $P_a\leftrightarrow R_a$ and the role of the fields $e^a\leftrightarrow h^a$. Preliminary inspection of the tables also with more elements does not seem to lead to the realization of the mixed situations where $\lambda_0\lambda_1=\lambda_3$ and $\lambda_0\lambda_3=\lambda_1$. All these types of the enlargements are justified within the $S$-expansion framework, but might be better to leave it for more detailed analysis in the future, as it goes beyond the use of the monoid requirement used in this work.

\section{Gauge transformations, curvatures and gravity actions}

In this section we focus on the gravity actions based on the found symmetries. To this end, we provide the necessary constituents and prescriptions concerning the extended field content, curvatures, transformations and invariant tensors. In general, any type of the shown here resonant algebras can be used in the arbitrary dimension. Looking at the types of generators $\{J_{ab,(i)},P_{a,(j)}\}=\{J_{ab},P_a,Z_{ab},R_a,...\}$, depending on the parity of $m$, there will be either the same amount of $J$-like and $P$-like generators or there will be one more kind of the former. Thus, being careful with the range of indices, we introduce separate $i,p=0,1,...$, and $j,q=0,1,...$ to take this into account. For a chosen algebra one builds the algebra-valued one-form connection,
\begin{equation}
A=\frac{1}{2}\omega ^{ab,(i)}{J}_{ab,(i)}+\frac{1}{\ell }e^{a,(j)}P_{a,(j)}\,,
\end{equation}
where group indices $a,b$ are running from $1$ to the spacetime dimension $D$. Then the zero-form gauge parameter,
\begin{equation}
\Theta =\frac{1}{2}\,\Lambda ^{ab,(i)}{J}_{ab,(i)}+\xi ^{a,(j)}P_{a,(j)}\,,
\end{equation}%
defines the infinitesimal gauge transformation according to
\begin{align}
\delta _{\Theta }\omega ^{ab,(i)}{J}_{ab,(i)} &=-\left( d\Lambda ^{ab,(i)}{J%
}_{ab,(i)}+(\omega ^{ac,(i)}\Lambda _{c}^{\;b,(p)}+\omega ^{bc,(i)}\Lambda
_{\;c}^{a,(p)})\lambda_{2i}\lambda_{2p}\tilde{J}_{ab}\right) \nonumber \\
&-\frac{1}{\ell }(e^{a,(j)}\xi ^{b,(q)}-e^{b,(j)}\xi
^{a,(q)})\lambda_{2j+1}\lambda_{2q+1}\tilde{J}_{ab}\,,\\
\delta _{\Theta }\frac{1}{\ell }e^{a,(j)}P_{a,(j)} &=-\left( d\xi
^{a,(j)}P_{a,(j)}+\omega _{\ b}^{a\ ,(i)}\xi ^{b,(j)}\lambda_{2i}\lambda_{2j+1}\tilde{P}%
_{a}\right)  \nonumber\\
&+\Lambda _{\ b}^{a\ ,(i)}\frac{1}{\ell }e^{b,(j)}\lambda_{2i}\lambda_{2j+1}\tilde{P}%
_{a}\,.
\end{align}
To obtain the final form, it is necessary to choose a particular semigroup multiplication law. Identity element $\lambda_0$ reproduces the standard transformation part, but it gets a new contribution every time $\lambda_0$ will again appear as a result of the multiplication between some other elements
\begin{align}
\delta _{\Theta }\omega _{\mu }^{ab}& =-D_{\mu }^{\omega }\Lambda ^{ab}-%
\frac{1}{\ell }(e_{\mu }^{a}\,\xi ^{b}-e_{\mu }^{b}\,\xi ^{a})+... \,, \notag \\
\frac{1}{\ell }\delta _{\Theta }e_{\mu }^{a}& =-D_{\mu }^{\omega }\xi ^{a}+%
\frac{1}{\ell }\Lambda _{\;b}^{a}\,e_{\mu }^{b}\,+... \,.
\end{align}
Naturally, the transformations of other fields become even more rearranged going from one algebra to another. The same happens for the curvature two-form
\begin{align}
F &=\frac{1}{2}F^{ab,(i)}{J}_{ab,(i)}+\frac{1}{\ell }%
F^{a,(j)}P_{a,(j)}=\frac{1}{2}F^{ab,(i)}\lambda_{2i}\tilde{J}_{ab}+\frac{1}{\ell }%
F^{a,(j)}\lambda_{2j+1}\tilde{P}_{a}\,, 
\end{align}
which will be resulting from
\begin{align}
F &=\frac{1}{2}\left( d\omega ^{ab,(i)}\lambda_{2i}+\omega
^{ac,(i)}\omega _{c}^{\;b,(p)}\lambda_{2i}\lambda_{2p}+\frac{1}{\ell ^{2}}%
e^{a,(j)}e^{b,(q)}\lambda_{2j+1}\lambda_{2q+1}\right)\tilde{J}_{ab} \nonumber \\
&+\frac{1}{\ell }\left( de^{a,(j)}\lambda_{2j+1}+\omega _{\ b}^{a\
,(i)}e^{b,(j)}\lambda_{2i}\lambda_{2j+1}\right)\tilde{P}_{a}\,.
\end{align}

One could try to construct an action by hand to get full invariance or give away some part of the symmetries, for example keeping only the Lorentz invariance like in Ref.~\cite{deAzcarraga:2010sw}. Deformed BF theory for $4D$ \cite{Durka:2011nf,Durka:2012wd} offers nice control on encoding the particular invariance through the additional auxiliary two-forms. In odd dimensions the Chern-Simons theory (CS) \cite{Chamseddine:1989nu} assures by construction the action invariance due to the full local gauge symmetry (local Lorentz and AdS boosts). In even dimensions one needs to use the Born-Infeld (BI) action with the invariance due to the Lorentzian subalgebra \cite{Concha:2013uhq,Concha:2014vka}. The remarkable advantage of the $S$-expansion method lies in the fact that it automatically provides the invariant tensor for the $S$-expanded algebra $\mathfrak{G}= S \times \mathfrak{g}$ in terms of the original $\mathfrak{g}=AdS$ tensor. For the $(2n+1)$-dimensional expanded Chern-Simons gravity action \cite{Izaurieta:2006zz} it is defined as follows
\begin{align}
\left\langle J_{a_{1}a_{2},\left(  k_{1}\right)  }\cdots J_{a_{2n-1}%
a_{2n},\left(  k_{n}\right)  }P_{a_{2n+1},\left(  k_{n+1}\right)
}\right\rangle &=\sigma_{2j+1}\delta_{k\left(  k_{1}%
,k_{2},\ldots,k_{n+1}\right)  }^{j}\frac{2^{n}}{n+1}\epsilon_{a_{1}a_{2}\ldots a_{2n+1}}\,.
\end{align}
An arbitrary dimensionless constant $\sigma_\alpha$, depending on the algebra, through its index introduces specific components of the invariant tensor. Effectively, for a chosen set of generators, index $\alpha$ in
\begin{align}
\left\langle (\lambda_{2k_1}\cdots\lambda_{2k_n}\lambda_{2k_{n+1}+1})\tilde{J}_{a_1 a_2}\cdots\tilde{J}_{a_{2n-1}a_{2n}}\tilde{P}_{a_{2n+1}}\right\rangle &=\sigma_{\alpha}\frac{2^n}{n+1} \epsilon_{a_1 a_2 \ldots a_{2n+1}}\,,
\end{align}
will be determined simply as an outcome of the multiplication of the related $\lambda$'s
\begin{align}
\lambda_{2k_1}\cdots\lambda_{2k_n}\lambda_{2k_{n+1}+1}=\lambda_{\alpha}\,.
\end{align}
Obviously, the presence of $0_S$ acting on the generators causes the whole $\left\langle\cdots\right\rangle=0$. 

Similarly, in even $2n$-dimensions an invariant tensor for the Born-Infeld action \cite{Concha:2013uhq} is given~as
\begin{align}
\left\langle (\lambda_{2k_1}\cdots\lambda_{2k_n})\tilde{J}_{a_1 a_2}\cdots\tilde{J}_{a_{2n-1}a_{2n}}\right\rangle &=\sigma_{\alpha}\frac{2^{n-1}}{n} \epsilon_{a_1 a_2 \ldots a_{2n}}\,.
\end{align}
The $S$-expansion concerns now the Lorentz algebra $\mathfrak{so}(2n-1,1)$ and uses as the semigroup $\bar{S}_0=\{0_S,\lambda_0,\lambda_2,...\}$. In that case an invariance is achieved only for the subalgebra $\mathfrak{L}=\bar{S}_0\times V_0$.

Both BI and CS prescriptions were thoroughly analyzed in various dimensions and for many different algebras. Let us now, without going much into details concerning the factors or constants, just shortly highlight the main features of the $5D$ CS and $4D$ BI actions based on all algebras with  $m=4\text{ and }5$.  Focusing only on the purely gravity content $(\omega,e$), the final CS terms depend on the chosen algebra and belong to the following sectors of the invariant tensor:
\begin{equation*}
\arrayrulecolor{Grey!50}%
\begin{tabular}{|c|c|c|c|c|c|c|c|c|c|}
\hline
~$5D$ CS~ & \multicolumn{1}{c|}{~~Poincar\'{e}-like~~} & \multicolumn{2}{|c|}{
~Maxwell~} & ~AdS-like~ & \multicolumn{5}{|c|}{~Maxwell-like~} \\ \hline
terms & $m=4$ and $5$ & ~~$\mathfrak{B}_{4}$~~ & ~~$\mathfrak{C}_{4}$~~ & $%
m=5$ & ~~$\mathfrak{B}_{5}$~~ & ~~$\mathfrak{C}_{5}$~~ & ~~$\mathfrak{D}_{5}$%
~~ & ~tables from (\ref{Max-like-2})~ & ~rest with $0_{s}$~ \\ \hline
$RRe$ & $\sigma _{1}$ & $\sigma _{1}$ & $\sigma _{1}$ & $\sigma _{1}$ & $%
\sigma _{1}$ & $\sigma _{1}$ & $\sigma _{1}$ & $\sigma _{1}$ & $\sigma _{1}$
\\ \hline
$Re^{3}$ & $0$ & $0$ & $\sigma _{1}$ & $\sigma _{1}$ & $\sigma _{3}$ & $%
\sigma _{3}$ & $\sigma _{3}$ & $\sigma _{1}$ & $0$ \\ \hline
$e^{5}$ & $0$ & $0$ & $\sigma _{1}$ & $\sigma _{1}$ & $0$ & $\sigma _{1}$ & $%
\sigma _{3}$ & $\sigma _{1}$ & $0$ \\ \hline
\end{tabular}%
\ \arrayrulecolor{White}
\end{equation*}
Extra field content naturally comes with its own set of $\sigma_\alpha$ constants. The most non-trivial shift of the various terms into different invariant tensor components happens for the Maxwell $\mathfrak{B}_5,\mathfrak{C}_5,\mathfrak{D}_5$ algebras. This was used in Refs.~\cite{Edelstein:2006se,Izaurieta:2009hz} as a starting point to find the Einstein-Hilbert term as a special limit of the CS theory and recently further explored in Ref.~\cite{Concha:2016kdz} in the context of the Pure Lovelock theory. 

Looking at an analogous table for the purely gravitational terms for $4D$ Born-Infeld actions we notice similar relations:
\begin{equation*}
\arrayrulecolor{Grey!50}%
\begin{tabular}{|c|c|c|c|c|c|c|c|c|c|}
\hline
~$4D$ BI~ & \multicolumn{1}{c|}{~~Poincar\'{e}-like~~} & \multicolumn{2}{|c|}{
~Maxwell~} & ~AdS-like~ & \multicolumn{5}{|c|}{Maxwell-like} \\ \hline
terms & ~$m=4\ $and $5$~ & ~~$\mathfrak{B}_{4}$~~ & ~~$\mathfrak{C}_{4}$~~ & 
~~$m=5$~~ & ~~$\mathfrak{B}_{5}$~~ & ~~$\mathfrak{C}_{5}$~~ & ~~$\mathfrak{D}%
_{5}$~~ & ~tables from (\ref{Max-like-2})~ & ~rest with $0_{s}$~ \\ \hline
$RR$ & $\sigma _{0}$ & $\sigma _{0}$ & $\sigma _{0}$ & $\sigma _{0}$ & $%
\sigma _{0}$ & $\sigma _{0}$ & $\sigma _{0}$ & $\sigma _{0}$ & $\sigma _{0}$
\\ \hline
$Re^{2}$ & $0$ & $\sigma _{2}$ & $\sigma _{2}$ & $\sigma _{0}$ & $\sigma _{2}
$ & $\sigma _{2}$ & $\sigma _{2}$ & $\sigma _{2}$ & $\sigma _{2}$ \\ \hline
$e^{4}$ & $0$ & $0$ & $\sigma _{2}$ & $\sigma _{0}$ & $0$ & $\sigma _{0}$ & $%
\sigma _{2}$ & $\sigma _{2}$ & $0$ \\ \hline
\end{tabular}%
\ \arrayrulecolor{White}
\end{equation*}

The full construction of the $4D$ BI actions based on $\mathfrak{B}_4$ and $\mathfrak{C}_4$ was given in details in Ref.~\cite{Salgado:2014qqa}. Besides repeating the same procedure for the other cases, it would be especially interesting to use $\tilde{B}_4$ and $C_4$ algebras from (\ref{Poincare-like-m4}) and mimic the Einstein-Hilbert action with the cosmological constant term through the additional constraint $D^\omega k^{ab}+k^{ac}\wedge k_{c}^{~b}= \frac{1}{\ell^2} e^a \wedge e^b$. 

For purely gravity content it could be regarded as a generic behavior, not only for $\mathfrak{C}_5$ in $5D$ but more generally, that the $\mathfrak{C}_D$ algebra in any odd $D\geq 5$ dimensions puts into the same invariant sector the maximal (dimensionally continued Euler) and minimal (cosmological constant) CS terms. Similarly, the requirement of vanishing the cosmological constant term from the CS Lagrangian but preserving all the others implies using the $\mathfrak{B}_m$ family. Analogously, we can show the same for the BI maximal and minimal terms for $\mathfrak{C}_5$ in $4D$, and more generally $\mathfrak{C}_{D+1}$ algebra in arbitrary even $D\geq 4$ dimensions. For the CS/BI limit an unique characteristic of the $\mathfrak{D}_m$ family could lead to the Einstein-Hilbert with $\Lambda$ term, but only at the level of action. It would be done by switching off the extra fields and forcing the part of the invariant tensor associated with maximal term to vanish by setting particular dimensionless constants $\sigma_\alpha$ to zero. 

However, the case of Pure Lovelock gravity teaches that the same sectors are not the whole story: sometimes it is necessary to obtain relative sign difference between the terms to admit proper vacuum and solutions. Nevertheless, for a sufficient amount of different types of the generators, thus high enough $m$, one can ultimately assure that each of the terms we are interested in could come with a separate arbitrary constant. That allows us to explicitly encode any relation between them. Mind that still this kind of immersion is not totally unrestricted. For instance, although $\mathfrak{C}_7$ finally assured in Ref.~\cite{Concha:2016kdz} the desired PL action limit from $5D$ CS, it still failed with a dynamical limit. Even though we can demand that the extra fields vanish, their variations bring the additional highly non-trivial conditions on the $\omega,e$ terms, and further special identification among the extra fields are required to overcome this issue. Maybe use of the other algebra offering different extra field configuration could help to improve that result.

In the $4D$ non-geometrical construction, which uses the Hodge $*$ star operator, it is possible to introduce $R(\omega+k)\wedge *R(\omega+k)=F(\bar{\omega})\wedge*F(\bar{\omega})$ term motivated by the direct sum algebras $C_4\equiv ISO\oplus Lorentz$ and $\mathfrak{C}_4\equiv AdS\oplus Lorentz$ (see Refs.~\cite{Durka:2011va, deAzcarraga:2012zv} and Appendix \ref{AppB}). The final action can then have the form of GR (related to first piece of algebra: Poincar\'{e} or AdS) coupled with the Yang-Mills term for the gauge group related to the second piece of algebra. Such an action is invariant due to the transformations generated by both Lorentz-like generators $L_{ab}=Z_{ab}$ and $N_{ab}=J_{ab}-Z_{ab}$. Given in Ref.~\cite{Durka:2011va} a possible link to the biconnection/bitetrad/bimetric theories could be now studied in the much more general framework of other algebras. 

Detailed discussion concerning all the mentioned applications will be left to the future work.

\section{Summary}
Imposing specific conditions on the wide class of the $S$-expanded resonant algebras, introduced in Refs.~\cite{Caroca:2011qs,Andrianopoli:2013ooa}, not only assured consistent grounds for many valuable and interesting algebras, but it also limited the overwhelming vastness of an algebraic examples. We find that expansions for the abelian semigroups with the zero and identity, simultaneously obeying the resonant decomposition, are well defined and straightforwardly applicable in the gravity context.

Among the results we have found all known Maxwell families of types $\mathfrak{B},\mathfrak{C},\mathfrak{D}$ and managed to generalize their description to the new examples. New enlargements containing the \textit{Poincar\'{e}} and $AdS$ as the subalgebras put it all in a greater perspective, which in the end allows for a complete generalization of the bosonic enlargements with an arbitrary number of the Lorentz-like and translational-like generators.

The wide class of algebras derived here delivers non-trivial extension to the construction of gauge gravity theories. Certainly, there is still a lot of work required to find a satisfactory interpretation of the extra field content, a description of the new symmetries and understand their consequences. The motivation is not very different from supersymmetry - to extend the general notion to accommodate extra symmetries/fields and go beyond the starting setup. Some applications related to the dark energy/matter and cosmological constant term were considered but it would be very interesting to look also at the supersymmetric versions of the presented here algebras. Some of them were already analyzed in Refs.~\cite{Izaurieta:2006zz,deAzcarraga:2002xi,Izaurieta:2006aj,Concha:2014xfa,Concha:2014tca}. New examples could help us even better understand underlying relations between various supergravity theories.

\section*{Acknowledgment}
Many thanks to Bogumi\l{} Kosza\l{}ka for providing the tool for checking the associativity and generating all the resonant tables going beyond the Maxwell algebras. One can find these materials supplemented on the webpage \href{https://resonantalgebras.wordpress.com/}{resonantalgebras.wordpress.com}.

Additional thanks to Marcin Kosz\'{o}w for a discussion concerning uniform multiplication law for the Maxwell algebras, as well as to N.~Merino, P.~Concha, E.~Rodr\'{\i}guez, C.~Inostroza, and F.~Izaurieta for the valuable comments and introduction to a subject of the $S$-expansion.

This work was supported by the Chilean FONDECYT Project No. 3140267.

\appendix
\section{Algebra from the multiplication table}
\label{AppA}
For a given semigroup multiplication table it is possible to instantly read off the corresponding Lie algebra. The expanded generators can be schematically collected in a similar commutation table, of course keeping in mind particular structure constants belonging to the commutators of $\left[X_{..},X_{..}\right]$, $\left[X_{..},X_{.}\right]$ and $\left[X_{.},X_{.}\right]$. Taking as an example the original Maxwell algebra, where $0_S\mathbb{T}_M=0$, we see that
\\[-1.5em]
\begin{align}
\footnotesize
\begin{squarecells}{4}
$\mathfrak{B}_4$  &     $\lambda_0$ & $\lambda_1$     & $\lambda_2$     \nl
$\lambda_0$ & \kx $\lambda_0$ & \ky $\lambda_1$ & \kw $\lambda_2$ \nl
$\lambda_1$ & \ky $\lambda_1$ & \kw $\lambda_2$ & \ko $0_S$ \nl
$\lambda_2$ & \kw $\lambda_2$ & \ko $0_S$ & \ko $0_S$ \nl 
\end{squarecells}
\qquad
semigroup \leftrightarrow algebra
\qquad
\begin{squarecells}{4}
 $\left[~,~\right]$ &      $J_{..}$ & $P_{.}$     & $Z_{..}$     \nl
$J_{..}$ & \kx $J_{..}$ & \ky $P_{.}$ & \kw $Z_{..}$ \nl
$P_{.}$ & \ky $P_{.}$ & \kw $Z_{..}$ & \ko $0$ \nl
$Z_{..}$ & \kw $Z_{..}$ & \ko $0$ & \ko $0$ \nl 
\end{squarecells}
\end{align}
with the generators defined as
\begin{align}
J_{ab}=\lambda_0\tilde{J}_{ab}\,,\qquad P_{a}=\lambda_1\tilde{P}_a\,,\qquad Z_{ab}=\lambda_2\tilde{J}_{ab}\,.
\end{align}
Commutation relations can be directly read off from the table on the right, or explicitly derived using the original AdS algebra given in Eqs.~(\ref{equ1})-(\ref{equ3}),
\begin{align}
\left[  J_{ab,}J_{cd}\right]   &  =\lambda_0\lambda_0(\eta_{bc}\tilde{J}_{ad}-\eta_{ac}\tilde{J}_{bd}+\eta
_{ad}\tilde{J}_{bc}-\eta_{bd}\tilde{J}_{ac})=\eta_{bc}J_{ad}-\eta_{ac}J_{bd}+\eta
_{ad}J_{bc}-\eta_{bd}J_{ac}\,,\nonumber\\
\left[  J_{ab},Z_{cd}\right]   &  =\lambda_0\lambda_2(\eta_{bc}\tilde{J}_{ad}-\eta_{ac}\tilde{J}_{bd}+\eta
_{ad}\tilde{J}_{bc}-\eta_{bd}\tilde{J}_{ac})=\eta_{bc}Z_{ad}-\eta_{ac}Z_{bd}+\eta
_{ad}Z_{bc}-\eta_{bd}Z_{ac}\,,\nonumber\\
\left[  Z_{ab,}Z_{cd}\right]   &  =\lambda_2\lambda_2(\eta_{bc}\tilde{J}_{ad}-\eta_{ac}\tilde{J}_{bd}+\eta
_{ad}\tilde{J}_{bc}-\eta_{bd}\tilde{J}_{ac})=0\,,\nonumber\\
\left[  J_{ab}%
,P_{c}\right] & =\lambda_0\lambda_1(\eta_{bc}\tilde{P}_{a}-\eta_{ac}\tilde{P}_{b})=\eta_{bc}P_{a}-\eta_{ac}P_{b}\,,\nonumber\\
\left[  Z_{ab}%
,P_{c}\right] & =\lambda_2\lambda_1(\eta_{bc}\tilde{P}_{a}-\eta_{ac}\tilde{P}_{b})=0\,,\nonumber\\
\left[  P_{a},P_{b}\right]   & =\lambda_1\lambda_1\tilde{J}_{ab}=Z_{ab}\,.
\end{align}

\section{Generalized change of basis}
\label{AppB}
The generalized change of basis presented in Ref.~\cite{Concha:2016hbt} originates from the result of Ref.~\cite{Soroka:2006aj}, where generators are reorganized to bring together $L_{ab}=Z_{ab}$ and $L_a=P_{a}$, and to introduce a shift in the definition of $N_{ab}=J_{ab}-Z_{ab}$. Then, the two algebra tables labeled by the subindex $m=4$, being Poincar\'{e}-like $C_4$  and Maxwell $\mathfrak{C}_4$,
\\[-1.5em]
\begin{align}
\footnotesize
\begin{squarecells}{4}
 $C_4$ &      $J_{..}$ & $P_{.}$     & $Z_{..}$     \nl
$J_{..}$ & \kx $J_{..}$ & \ky $ P_{.}$ & \kw $ Z_{..}$ \nl
$P_{.}$ & \ky $ P_{.}$ & \ko $0$ & \ky $ P_{.}$ \nl
$Z_{..}$ & \kw $Z_{..}$ & \ky $ P_{.}$ & \kw $ Z_{..}$ \nl 
\end{squarecells}
\qquad
\begin{squarecells}{4}
 $\mathfrak{C}_4$ &      $J_{..}$ & $P_{.}$     & $Z_{..}$     \nl
$J_{..}$ & \kx $J_{..}$ & \ky $P_{.}$ & \kw $Z_{..}$ \nl
$P_{.}$ & \ky $P_{.}$ & \kw $Z_{..}$ & \ky $P_{.}$ \nl
$Z_{..}$ & \kw $Z_{..}$ & \ky $P_{.}$ & \kw $Z_{..}$ \nl 
\end{squarecells}
\end{align}
by the means of $[N_{..},L_{..}]=0$ and $[N_{..},L_{.}]=0$, happen to be equivalent to

\begin{align}
\footnotesize
\begin{squarecells}{4}
 $C_4$ &      $L_{..}$ & $L_{.}$     & $N_{..}$     \nl
$L_{..}$ & \ku $L_{..}$ & \ku $L_{.}$ & \ko $0$ \nl
$L_{.}$ & \ku $L_{.}$ & \ko $0$ & \ko $0$ \nl
$N_{..}$ & \ko $0$ & \ko $0$ & \ku $N_{..}$ \nl 
\end{squarecells}
\qquad
\begin{squarecells}{4}
$\mathfrak{C}_4$ &      $L_{..}$ & $L_{.}$     & $N_{..}$     \nl
$L_{..}$ & \ku $L_{..}$ & \ku $L_{.}$ & \ko $0$ \nl
$L_{.}$ & \ku $L_{.}$ & \ku $L_{..}$ & \ko $0$ \nl
$N_{..}$ & \ko $0$ & \ko $0$ & \ku $N_{..}$ \nl 
\end{squarecells}
\end{align}
obviously forming the direct sums of subalgebras: $\textit{Poincar\'{e}}\oplus Lorentz$ and $AdS\oplus Lorentz$.\\

\noindent The same can be repeated for the case with $m=5$. Indeed, three particular algebras 
\\[-1.5em]
\begin{align}
\footnotesize
\begin{squarecells}{5}
$D_5$  &     $J_{..}$ & $P_{.}$     & $Z_{..}$    & $R_{.}$ \nl
$J_{..}$ & \kx $J_{..}$ & \ky $P_{.}$ & \kw $Z_{..}$ & \kv $R_{.}$ \nl
$P_{.}$ & \ky $P_{.}$ & \ko $0$ & \kv $R_{.}$ & \ko $0$ \nl
$Z_{..}$ & \kw $Z_{..}$ & \kv $R_{.}$ & \kw $Z_{..}$ & \kv $R_{.}$ \nl
$R_{.}$ & \kv $R_{.}$ & \ko $0$ & \kv $R_{.}$ & \ko $0$ \nl
\end{squarecells}
\quad
\begin{squarecells}{5}
$\mathcal{D}_5$  &     $J_{..}$ & $P_{.}$     & $Z_{..}$    & $R_{.}$ \nl
$J_{..}$ & \kx $J_{..}$ & \ky $P_{.}$ & \kw $Z_{..}$ & \kv $R_{.}$ \nl
$P_{.}$ & \ky $P_{.}$ & \kx $J_{..}$ & \kv $R_{.}$ & \kw $Z_{..}$ \nl
$Z_{..}$ & \kw $Z_{..}$ & \kv $R_{.}$ & \kw $Z_{..}$ & \kv $R_{.}$ \nl 
$R_{.}$ & \kv $R_{.}$ & \kw $Z_{..}$ & \kv $R_{.}$ & \kw $Z_{..}$ \nl
\end{squarecells}
\quad
\begin{squarecells}{5}
$\mathfrak{D}_5$  &     $J_{..}$ & $P_{.}$     & $Z_{..}$    & $R_{.}$ \nl
$J_{..}$ & \kx $J_{..}$ & \ky $P_{.}$ & \kw $Z_{..}$ & \kv $R_{.}$ \nl
$P_{.}$ & \ky $P_{.}$ & \kw $Z_{..}$ & \kv $R_{.}$ & \kw $Z_{..}$ \nl
$Z_{..}$ & \kw $Z_{..}$ & \kv $R_{.}$ & \kw $Z_{..}$ & \kv $R_{.}$ \nl 
$R_{.}$ & \kv $R_{.}$ & \kw $Z_{..}$ & \kv $R_{.}$ & \kw $Z_{..}$ \nl
\end{squarecells}
\end{align}
with the redefinitions $L_{ab}=Z_{ab}$ and $L_a=R_{a}$ along with  $N_{ab}=J_{ab}-Z_{ab}$ and $N_{a}=P_a-R_a$ are equivalent to
\\[-1.5em]
\begin{align}
\footnotesize
\begin{squarecells}{5}
$D_5$  &     $L_{..}$ & $L_{.}$     & $N_{..}$    & $N_{.}$ \nl
$L_{..}$ & \ku $L_{..}$ & \ku $L_{.}$ & \ko $0$ & \ko $0$ \nl
$L_{.}$ & \ku $L_{.}$ & \ko $0$ & \ko $0$ & \ko $0$ \nl
$N_{..}$ & \ko $0$ & \ko $0$ & \ku $N_{..}$ & \ku $N_{.}$ \nl 
$N_{.}$ & \ko $0$ & \ko $0$ & \ku $N_{.}$ & \ko $0$ \nl
\end{squarecells}
\quad
\begin{squarecells}{5}
$\mathcal{D}_5$  &     $L_{..}$ & $L_{.}$     & $N_{..}$    & $N_{.}$ \nl
$L_{..}$ & \ku $L_{..}$ & \ku $L_{.}$ & \ko $0$ & \ko $0$ \nl
$L_{.}$ & \ku $L_{.}$ & \ku $L_{..}$ & \ko $0$ & \ko $0$ \nl
$N_{..}$ & \ko $0$ & \ko $0$ & \ku $N_{..}$ & \ku $N_{.}$ \nl 
$N_{.}$ & \ko $0$ & \ko $0$ & \ku $N_{.}$ & \ku $N_{..}$ \nl
\end{squarecells}
\quad
\begin{squarecells}{5}
$\mathfrak{D}_5$  &     $L_{..}$ & $L_{.}$     & $N_{..}$    & $N_{.}$ \nl
$L_{..}$ & \ku $L_{..}$ & \ku $L_{.}$ & \ko $0$ & \ko $0$ \nl
$L_{.}$ & \ku $L_{.}$ & \ku $L_{..}$ & \ko $0$ & \ko $0$ \nl
$N_{..}$ & \ko $0$ & \ko $0$ & \ku $N_{..}$ & \ku $N_{.}$ \nl 
$N_{.}$ & \ko $0$ & \ko $0$ & \ku $N_{.}$ & \ko $0$ \nl
\end{squarecells}
\end{align}
clearly forming the direct sums of $\textit{Poincar\'{e}}\oplus \textit{Poincar\'{e}}$ and $AdS\oplus AdS$ and $AdS\oplus \textit{Poincar\'{e}}$. 
The explicit prescription for the change of basis in a general case can be found in Ref.~\cite{Concha:2016hbt}. 

Mind that in the algebraic sums above one deals with $L_{ab}$ and $N_{ab}$, which represent a peculiar kind of the Lorentz generator that commutes with other generators. This is very different from what happens to the standard  $J_{ab}$ outcomes. It leave us with a question if these Lorentz generators could be physical or they are just a mere artifact of the mathematical description. From the formal point of view, introducing an identity element might not exhaust all the possibilities, as the semigroup equivalents of the algebraic redefinitions above, along with other examples containing the zero elements (see \textit{resonantalgebras.wordpress.com}), still seem to represent valid algebras with possible use in the gravity models.


\begin{thebibliography}{99}     
\small

\bibitem{Concha:2016hbt} 
P.~K.~Concha, R.~Durka, N.~Merino and E.~K.~Rodríguez,
``New family of Maxwell like algebras,''
Phys.\ Lett.\ B {\bf 759}, 507 (2016)
doi:10.1016/j.physletb.2016.06.016
[arXiv:1601.06443 [hep-th]].

\bibitem{Izaurieta:2006zz} 
  F.~Izaurieta, E.~Rodr\'{\i}guez and P.~Salgado,
  ``Expanding Lie (super)algebras through Abelian semigroups,''
  J.\ Math.\ Phys.\  {\bf 47}, 123512 (2006)
    [hep-th/0606215].
  
\bibitem{Salgado:2014qqa} 
  P.~Salgado and S.~Salgado,
  ``$\mathfrak{so}(D-1,1)\otimes \mathfrak{so}(D-1,2)$ algebras and gravity,''
  Phys.\ Lett.\ B {\bf 728}, 5 (2014).
 
\bibitem{Diaz:2012zza} 
  J.~Diaz, O.~Fierro, F.~Izaurieta, N.~Merino, E.~Rodr\'{i}guez, P.~Salgado and O.~Valdivia,
  ``A generalized action for (2 + 1)-dimensional Chern-Simons gravity,''
  J.\ Phys.\ A {\bf 45}, 255207 (2012)
  [arXiv:1311.2215 [gr-qc]].
  
\bibitem{Schrader:1972zd} 
  R.~Schrader,
  ``The maxwell group and the quantum theory of particles in classical homogeneous electromagnetic fields,''
  Fortsch.\ Phys.\  {\bf 20}, 701 (1972).

\bibitem{Bacry:1970ye} 
  H.~Bacry, P.~Combe and J.~L.~Richard,
  ``Group-theoretical analysis of elementary particles in an external electromagnetic field. 1. the relativistic particle in a constant and uniform field,''
  Nuovo Cim.\ A {\bf 67}, 267 (1970).
                          
\bibitem{Soroka:2004fj} 
  D.~V.~Soroka and V.~A.~Soroka,
  ``Tensor extension of the Poincar\'{e}' algebra,''
  Phys.\ Lett.\ B {\bf 607}, 302 (2005)
  [hep-th/0410012].
                
\bibitem{Soroka:2006aj} 
  D.~V.~Soroka and V.~A.~Soroka,
  ``Semi-simple extension of the (super)Poincar\'{e} algebra,''
  Adv.\ High Energy Phys.\  {\bf 2009}, 234147 (2009)
  [hep-th/0605251].
  
\bibitem{deAzcarraga:2002xi} 
  J.~A.~de Azcarraga, J.~M.~Izquierdo, M.~Picon and O.~Varela,
 ``Generating Lie and gauge free differential (super)algebras by expanding Maurer-Cartan forms and Chern-Simons supergravity,''
  Nucl.\ Phys.\ B {\bf 662}, 185 (2003)
  [hep-th/0212347].

\bibitem{Gomis:2009dm} 
  J.~Gomis, K.~Kamimura and J.~Lukierski,
  ``Deformations of Maxwell algebra and their Dynamical Realizations,''
  JHEP {\bf 0908}, 039 (2009)
  [arXiv:0906.4464 [hep-th]].
  
  \bibitem{Durka:2011nf} 
   R.~Durka, J.~Kowalski-Glikman and M.~Szczachor,
  ``Gauged AdS-Maxwell algebra and gravity,''
    Mod.\ Phys.\ Lett.\ A {\bf 26}, 2689 (2011)
   [arXiv:1107.4728 [hep-th]].
      
\bibitem{Durka:2012wd} 
  R.~Durka,
  ``Deformed BF theory as a theory of gravity and supergravity,''
  arXiv:1208.5185 [gr-qc].
 
\bibitem{deAzcarraga:2010sw} 
  J.~A.~de Azcarraga, K.~Kamimura and J.~Lukierski,
  ``Generalized cosmological term from Maxwell symmetries,''
  Phys.\ Rev.\ D {\bf 83}, 124036 (2011)
  [arXiv:1012.4402 [hep-th]].
  
\bibitem{Durka:2011va} 
  R.~Durka and J.~Kowalski-Glikman,
  ``Local Maxwell symmetry and gravity,''
  arXiv:1110.6812.
  
\bibitem{Inostroza:2014vua} 
  C.~Inostroza, A.~Salazar and P.~Salgado,
  ``Brans–Dicke gravity theory from topological gravity,''
  Phys.\ Lett.\ B {\bf 734}, 377 (2014).

  \bibitem{Salgado:2014jka} 
    P.~Salgado, R.~J.~Szabo and O.~Valdivia,
    ``Topological gravity and transgression holography,''
    Phys.\ Rev.\ D {\bf 89}, no. 8, 084077 (2014)
    doi:10.1103/PhysRevD.89.084077
    [arXiv:1401.3653 [hep-th]].
    
\bibitem{Durka:2011gm} 
  R.~Durka, J.~Kowalski-Glikman and M.~Szczachor,
  ``AdS-Maxwell superalgebra and supergravity,''
  Mod.\ Phys.\ Lett.\ A {\bf 27}, 1250023 (2012)
  [arXiv:1107.5731 [hep-th]].

\bibitem{Kamimura:2011mq} 
  K.~Kamimura and J.~Lukierski,
  ``Supersymmetrization Schemes of D=4 Maxwell Algebra,''
  Phys.\ Lett.\ B {\bf 707}, 292 (2012)
  [arXiv:1111.3598 [math-ph]].
  
\bibitem{deAzcarraga:2012zv} 
  J.~A.~de Azcarraga, J.~M.~Izquierdo, J.~Lukierski and M.~Woronowicz,
  ``Generalizations of Maxwell (super)algebras by the expansion method,''
  Nucl.\ Phys.\ B {\bf 869}, 303 (2013)
  [arXiv:1210.1117 [hep-th]].
                
\bibitem{Edelstein:2006se} 
  J.~D.~Edelstein, M.~Hassaine, R.~Troncoso and J.~Zanelli,
  ``Lie-algebra expansions, Chern-Simons theories and the Einstein-Hilbert Lagrangian,''
  Phys.\ Lett.\ B {\bf 640}, 278 (2006)\qquad[hep-th/0605174].

\bibitem{Izaurieta:2009hz} 
  F.~Izaurieta, E.~Rodr\'{\i}guez, P.~Minning, P.~Salgado and A.~Perez,
  ``Standard General Relativity from Chern-Simons Gravity,''
  Phys.\ Lett.\ B {\bf 678}, 213 (2009)
  [arXiv:0905.2187 [hep-th]].

\bibitem{Concha:2013uhq} 
  P.~K.~Concha, D.~M.~Pe\~{n}afiel, E.~K.~Rodr\'{\i}guez and P.~Salgado,
  ``Even-dimensional General Relativity from Born-Infeld gravity,''
  Phys.\ Lett.\ B {\bf 725}, 419 (2013)
  doi:10.1016/j.physletb.2013.07.019
  [arXiv:1309.0062 [hep-th]].
    
\bibitem{Concha:2014vka} 
  P.~K.~Concha, D.~M.~Pe\~{n}afiel, E.~K.~Rodr\'{\i}guez and P.~Salgado,
  ``Chern-Simons and Born-Infeld gravity theories and Maxwell algebras type,''
  Eur.\ Phys.\ J.\ C {\bf 74}, 2741 (2014)
  [arXiv:1402.0023 [hep-th]].

\bibitem{Concha:2016kdz} 
P.~K.~Concha, R.~Durka, C.~Inostroza, N.~Merino and E.~K.~Rodríguez,
``Pure Lovelock gravity and Chern-Simons theory,''
Phys.\ Rev.\ D {\bf 94}, no. 2, 024055 (2016)
doi:10.1103/PhysRevD.94.024055
[arXiv:1603.09424 [hep-th]].

\bibitem{Gonzalez:2014tta} 
  N.~Gonz\'{a}lez, P.~Salgado, G.~Rubio and S.~Salgado,
  ``Einstein–Hilbert action with cosmological term from Chern–Simons gravity,''
  J.\ Geom.\ Phys.\  {\bf 86}, 339 (2014).

\bibitem{Chamseddine:1989nu} 
  A.~H.~Chamseddine,
  ``Topological Gauge Theory of Gravity in Five-dimensions and All Odd Dimensions,''
  Phys.\ Lett.\ B {\bf 233}, 291 (1989).
  
\bibitem{Caroca:2011qs} 
  R.~Caroca, I.~Kondrashuk, N.~Merino and F.~Nadal,
  ``Bianchi spaces and their three-dimensional isometries as S-expansions of two-dimensional isometries,''
  J.\ Phys.\ A {\bf 46}, 225201 (2013)
  [arXiv:1104.3541 [math-ph]].

\bibitem{Andrianopoli:2013ooa} 
  L.~Andrianopoli, N.~Merino, F.~Nadal and M.~Trigiante,
  ``General properties of the expansion methods of Lie algebras,''
  J.\ Phys.\ A {\bf 46}, 365204 (2013)
  [arXiv:1308.4832 [gr-qc]].

\bibitem{Izaurieta:2006aj} 
  F.~Izaurieta, E.~Rodr\'{\i}guez and P.~Salgado,
  ``Eleven-dimensional gauge theory for the M algebra as an Abelian semigroup expansion of $osp(32|1)$,''
  Eur.\ Phys.\ J.\ C {\bf 54}, 675 (2008)
  [hep-th/0606225].
  
      \bibitem{Concha:2014xfa} 
      P.~K.~Concha and E.~K.~Rodr\'{\i}guez,
      ``Maxwell Superalgebras and Abelian Semigroup Expansion,''
      Nucl.\ Phys.\ B {\bf 886}, 1128 (2014)
      [arXiv:1405.1334 [hep-th]].
      
  \bibitem{Concha:2014tca} 
    P.~K.~Concha and E.~K.~Rodr\'{\i}guez,
    ``N = 1 Supergravity and Maxwell superalgebras,''
    JHEP {\bf 1409}, 090 (2014)
    [arXiv:1407.4635 [hep-th]].
  
\end{thebibliography}
\end{document}